\documentclass[jurnal]{IEEEtran}
\usepackage{amsfonts}
\usepackage{amssymb}
\usepackage{graphicx}
\usepackage{xcolor}
\usepackage{comment}

\usepackage{cite}
\usepackage{amsmath}
\usepackage{algorithmic}
\ifCLASSOPTIONcompsoc
  \usepackage[caption=false,font=normalsize,labelfont=sf,textfont=sf]{subfig}
\else
  \usepackage[caption=false,font=footnotesize]{subfig}
\fi

\usepackage{stfloats}
\begin{document}
%
\title{Unsupervised Sparse Unmixing of Atmospheric Trace Gases from Hyperspectral Satellite Data}
%
%
%

\author{Nicomino~Fiscante,~\IEEEmembership{Student Member,~IEEE,}
        Pia~Addabbo,~\IEEEmembership{Senior Member,~IEEE,}
        Filippo~Biondi,~\IEEEmembership{Member,~IEEE,}
		Gaetano~Giunta,~\IEEEmembership{Senior Member,~IEEE,}
		and~Danilo~Orlando,~\IEEEmembership{Senior Member,~IEEE}

\thanks{N. Fiscante and G. Giunta are with Department of Engineering, University of Roma TRE, 00154 Rome, Italy.}
\thanks{P. Addabbo is with Universit\`a degli Studi Giustino Fortunato, viale Raffale Delcogliano, 12, 82100 Benevento, Italy.}
\thanks{F. Biondi is with Italian Ministry of Defence.}
\thanks{D. Orlando is with Universit\`a degli Studi Niccol\`o Cusano, Via Don Carlo Gnocchi, 3, 00166 Roma, Italy.}

\thanks{Manuscript received ---; revised ---.}}

%
%

\markboth{Journal of \LaTeX\ Class Files,~Vol.~X, No.~Y, Month~Year}%
{Shell \MakeLowercase{\textit{et al.}}: Bare Demo of IEEEtran.cls for IEEE Journals}
%



\maketitle

\begin{abstract}
In this letter, a new approach for the retrieval of the vertical column concentrations of trace gases from hyperspectral satellite observations, is proposed. 
The main idea is to perform a linear spectral unmixing by estimating the abundances of trace gases spectral signatures in each mixed pixel collected by an imaging spectrometer in the ultraviolet region. 
To this aim, the sparse nature of the measurements is brought to light and the compressive sensing paradigm is applied to estimate the concentrations of the gases' endemembers given by an \textit{a priori} wide spectral library, including reference cross sections measured at different temperatures and pressures at the same time. 
The proposed approach has been experimentally assessed using both simulated and real hyperspectral dataset. Specifically, the experimental analysis relies on the retrieval of sulfur dioxide during volcanic emissions using data collected by the TROPOspheric Monitoring Instrument. To validate the procedure, we also compare the obtained results with the sulfur dioxide total column product based on the differential optical absorption spectroscopy technique and the retrieved concentrations estimated using the blind source separation.

\end{abstract}

\begin{IEEEkeywords}
Hyperspectral unmixing, Sparse data recovery, Trace gas concentration retrieval, TROPOspheric Monitoring Instrument, Unsupervised sparse learning.
\end{IEEEkeywords}

%
\IEEEpeerreviewmaketitle

\section{Introduction}
%
%
%
%
\IEEEPARstart{R}{ecent} research has clearly shown that hyperspectral satellite observations can be successfully used to map atmospheric trace gas concentrations throughout the planet. Retrieval algorithms play a primary role in this process, and several techniques have been developed in the last decade to improve the estimation of the atmospheric 
constituents, in terms of both accuracy and minimum detectable amounts. 

Differential Optical Absorption Spectroscopy (DOAS) is, perhaps, the most widely used technique for retrieving trace-gas abundances in the open atmosphere using the narrow-band absorption structures in the near UtraViolet (UV), VISible (VIS), and near-infrared wavelength regions \cite{book_DOAS}. It is also known that, in some applications, results from standard DOAS are affected by spectral interference from other gases, such as in sulfur dioxide (SO\textsubscript{2}) retrieval with a strong interference by the ozone (O$_3$) in the narrow $260-340$ nm vibrational band. To improve accuracy in this case, the Weighting Function DOAS (WFDOAS) has been developed with wavelength-dependent weighting functions used in place of the absorption cross sections \cite{acp_8_6137_2008}. Moreover, the main limitation of this technique is the need of \textit{a priori} information about the trace gas under analysis. Further investigations, aimed at improving the SO\textsubscript{2} retrievals, were developed using the principal component analysis to characterize the effects caused by both the physical processes and the measurement artifacts in the absence of SO\textsubscript{2} absorption \cite{LiCAN20136}. A different unmixing approach, based on Blind Source Separation (BSS), has been recently proposed in \cite{6071033,7163519}. This method replaces the constraint of perfect independence among sources with a minimum dependence criterion and represents an effective way for measuring the SO\textsubscript{2} from volcanic emission using the Ozone Monitoring Instrument (OMI) of National Aeronautics and Space Administration (NASA) \cite{6071033} and the nitrogen dioxide (NO\textsubscript{2}) from anthropogenic pollution using the SCanning Imaging Absorption spectroMeter for Atmospheric CHartographY (SCIAMACHY) of European Space Agency (ESA)\cite{7163519,7326687}.

In this contribution, we propose a different approach based on Sparse Learning Iterative Minimization (SLIM) \cite{5654598}, whose essence is the sparse unmixing of the spectral waveforms of individual trace gases, providing the estimation of their concentrations. This new approach is deeply different from DOAS, whose results, as any absorption spectroscopy method, are dependent from few selected reference cross sections used in the fitting \cite{book_DOAS,article_Richter}. On the contrary, the SLIM-based approach, thanks to the sparsity of the proposed model, including several cross sections taken at different temperatures and pressures, offers the possibility to automatically extract the best candidate trace gas present in the scene, realizing an unsupervised unmixing of the observation. Finally, to the best of the authors' knowledge, it is the first time that the sparse paradigm is used in this context as a tool to measure trace gas concentrations from satellite observations.


\section{Problem Statement}

Exploiting the Beer-Lambert law 
and according to the DOAS formulation \cite{book_DOAS}, we can deal with each pixel by means of the so-called linear mixing model, namely
		\begin{equation}
		\label{linear_mixing_model}
		z_m(\lambda)= \sum\limits_{n=1}^{N} \alpha_{m,n}{\sigma}_{n}(\lambda) + P_m(\lambda),
		\end{equation}
		where $\lambda$ is the wavelength, $z_m(\lambda)$, $m\in{1,\dots,M}$, is the $m-$th spectral pixel value, $N$ is the number of the total absorption gases, $\alpha_{m,n}$ and $\sigma_{n}(\lambda)$ are the positive column density abundance and the trace gases absorption cross section (endmembers) for the $n-$th species, respectively and, $P_m(\lambda)$ is a slowly varying polynomial accounting for multiple Rayleigh scattering, Mie scattering, and surface albedo \cite{OMI_NASA_IV, Rozanov_DOAS}. The slowly varying component of the spectral waveform can be efficiently removed by using a Savitzky-Golay filter, which provides a least-square smoothing of the noisy data, especially when the shape and amplitude of the waveform peaks must be preserved \cite{Ruffin_Savitzky-Golay}, thus, the output of such a filter has the following form
		\begin{equation}
		\label{IFOV_observation2}
		\tilde{z}_m(\lambda)= \sum\limits_{n=1}^{N} \alpha_{m,n}\tilde{\sigma}_{n}(\lambda) + n_m(\lambda),
		\end{equation}
		where $\tilde{\sigma}_{n}(\lambda)$ is the fast-varying component of ${\sigma}_{n}(\lambda)$ and $n_m(\lambda)$ accounts for the residual error. Now, since (\ref{IFOV_observation2}) is obtained for each spectral wavelength belonging to a given discrete finite set, i.e., $\{ \lambda_1,\ldots,\lambda_L \}$, we can use (\ref{IFOV_observation2}) to form the vector $\boldsymbol{z}_m$ given by
		\begin{equation}
		\label{IFOV_observation_vecto}
		\boldsymbol{z}_m = \sum\limits_{n=1}^{N} \alpha_{m,n}\boldsymbol{\sigma}_{n} + \boldsymbol{n}_m = \boldsymbol{S} \boldsymbol{\alpha}_m + \boldsymbol{n}_m,
		\end{equation}
		where $\boldsymbol{z}_m =\left[\tilde{z}_m(\lambda_1), \dots,\tilde{z}_m(\lambda_L)  \right]^T \in \mathbb{R}^{L \times 1}$ with $(\cdot)^T$ denoting the transpose, $\boldsymbol{\sigma}_{n} = \left[\tilde{\sigma}_{n}(\lambda_1),\dots,\tilde{\sigma}_{n}(\lambda_L) \right]^T\in \mathbb{R}^{L \times 1}$, $\boldsymbol{n}_m = \left[ n_m(\lambda_1),\dots,n_m(\lambda_L)  \right]^T \in \mathbb{R}^{L \times 1}$, $\boldsymbol{S} =[\boldsymbol{\sigma}_{1},\dots,\boldsymbol{\sigma}_{N}] \in \mathbb{R}^{L \times N}$ is the endmembers' matrix, i.e., the spectral library or dictionary, and $\boldsymbol{\alpha}_m = \left[ \alpha_{m,1}, \dots, \alpha_{m,N} \right]^T\in \mathbb{R}^{N \times 1}$ is the vector of the unknown abundances for the $m$th pixel. The error component $\boldsymbol{n}_m$, due principally to thermal and quantization noise, is modelled in term of an independent Gaussian process \cite{Kerekes2003HyperspectralIS, 4072455}, and, usually, it is also assumed to be spectrally uncorrelated \cite{4072455, Rasti_SURE}. This leads to the so called Gaussian mixture model which has been broadly used in hyperspectral unmixing since it considerably simplifies the unmixing process providing acceptable results \cite{TheilerFoySPIE2018,rs10030482}. Other models may also be considered for hyperspectral image \cite{5289080}, however, this linear model is the most widely used in the literature \cite{974727,6200362,5692827,6568877,7433396}.
		
		If we assume in (\ref{IFOV_observation_vecto}) that $\boldsymbol{S}$ contains all the available spectral endmembers also taken at different temperatures and pressures $(N \gg L)$, then $\boldsymbol{\alpha}_m$ becomes  a sparse vector.
		In fact, since the number of endmembers with a non-negligible absorption in the wavelengths' interval of interest is usually very small when compared with the size of the available spectral library, this means that most of the entries of $\boldsymbol{\alpha}_m$ are zero. 
		
		It is important to underline that the proposed estimation procedure based upon the above sparse model does not require any information about the availability of pure spectral signatures in the input data or that a certain endmember extraction algorithm preliminarily identifies such pure signatures. Specifically, the proposed procedure finds the optimal subset of signatures in the library that experience the most likely match with each mixed pixel in the scene. Thus, the procedure takes advantage of the availability of wide spectral libraries of the species leading to highly sparse data models and, hence, in the next section we develop an estimation procedure that takes advantage of this sparsity\footnote{The sparse model is fundamentally different from \cite{6071033,7163519,7326687}, where endmembers are \textit{a priori} selected as supposed to be present in the scene $(N < L)$.}.

\section{Trace Gas retrieval via Sparse Learning Iterative Minimization}

The main objective of the unmixing problem represented by (\ref{IFOV_observation_vecto}) consists of estimating the sparse vector $\boldsymbol{\alpha}_m$ assuming that endmembers' matrix is known.  To this end, we apply the SLIM framework \cite{5654598,8781902} assuming that $\boldsymbol{\alpha}_m$ is a random real vector ruled by a sparsity promoting prior \cite{gelmanbda04,10.5555/1051451}. It follows that
\begin{equation}
\boldsymbol{z}_m|\boldsymbol{\alpha}_m \sim \mathcal{N} (\boldsymbol{S} \boldsymbol{\alpha}_m,\boldsymbol{M}_m),
\label{bayesian_model}
\end{equation}
\noindent
whereas the prior of $\boldsymbol{\alpha}_m$ is\footnote{The sparsity promoting prior leads to a controlled and manageable sparse solution of the abundance vector. Different priors are possible \cite{gelmanbda04, 10.5555/1051451, Mohammad-Djafari,Akhtar_Naveed}.}
\begin{equation}
f(\boldsymbol{\alpha}_m;q) \propto  \prod_{i=1}^{N} e^{-\frac{1}{q}(\alpha_{m,i}^q-1)},
\label{bayesian_model}
\end{equation}
\noindent
where $q \in (0,1]$ is a parameter controlling the sparsity level. Supposing that $\boldsymbol{M}_m$ is known, or, it can be estimated from data, 
 we can estimate $\boldsymbol{\alpha}_m$ as follows
\begin{equation}
\max_{\boldsymbol{\alpha}_m}f(\boldsymbol{z}_m | \boldsymbol{\alpha}_m)f(\boldsymbol{\alpha}_m;q),
\label{MAP_rule}
\end{equation}
\noindent
where
\begin{equation}
f(\!\boldsymbol{z}_m | \boldsymbol{\alpha}_m\!)\!=\!\frac{(2\pi)^{-\frac{N}{2}}}{\sqrt{\text{det}(\boldsymbol{M}_m)}}e^{-\frac{1}{2}(\boldsymbol{z}_m\!-\!\boldsymbol{S}\boldsymbol{\alpha}_m)^{T}\!\boldsymbol{M}_m^{-1}\!(\boldsymbol{z}_m\!-\!\boldsymbol{S}\boldsymbol{\alpha}_m)},
\label{pdf_expression}
\end{equation}
is the probability density function of $\boldsymbol{z}_m$ given $\boldsymbol{\alpha}_m$ with $\text{det}(\cdot)$ denotes the determinant. By taking the negative logarithm, problem \eqref{MAP_rule} is equivalent to
\begin{equation}
\min_{\boldsymbol{\alpha}_m}\underbrace{\Bigg\{\frac{1}{2}\left\|\boldsymbol{y}_m-\boldsymbol{V} \boldsymbol{\alpha}_m\right\|^2_2 + \sum_{i=1}^{N} \frac{1}{q}(\alpha_{m,i}^q-1)\Bigg\}}_\text{$g_q(\boldsymbol{\alpha}_m)$},
\label{SLIM_estimation}
\end{equation}
\noindent
where $\boldsymbol{y}_m=\big ( \boldsymbol{M}_m \big )^{-1/2} \boldsymbol{z}_m$, $\boldsymbol{V}=\big ( \boldsymbol{M}_m \big )^{-1/2} \boldsymbol{S}$ and $|| \cdot ||_{2}$ denotes the Euclidean norm. Note that the first addendum of $g_q(\boldsymbol{\alpha}_m)$ corresponds to a fitting term, whereas the second term promotes the sparsity. 

Setting to zero the first derivative of $g_q(\boldsymbol{\alpha}_m)$ with respect to $\boldsymbol{\alpha}_m$ leads to:
\begin{equation}
\frac{d}{d\boldsymbol{\alpha}_m}[g_q(\boldsymbol{\alpha}_m)]=\boldsymbol{V}^{T}\boldsymbol{V}\boldsymbol{\alpha}_m-\boldsymbol{V}^{T}\boldsymbol{y}_m+\boldsymbol{P}^{-1}_q\boldsymbol{\alpha}_m=0,
\label{SLIM_derivate}
\end{equation}
\noindent
where $\boldsymbol{P}_q=$ diag$(\boldsymbol{p}_q)$, with $\boldsymbol{p}_q=[\alpha_{m,1}^{2-q} , \dots ,\alpha_{m,N}^{2-q}]^T$ and $\text{diag}(\cdot)$ denotes the diagonalization operation. Supposing that an initial estimate of $\boldsymbol{\alpha}_m$ is available, it is possible to apply a cyclic optimization procedure, and the step at the $n$th iteration can be expressed as\footnote{Negative values of the estimated vector entries are forced to zero.}
\begin{equation}
\boldsymbol{\alpha}^{(n)}_{m,q}=\boldsymbol{P}^{(n-1)}_q\boldsymbol{V}^{T}\left(\boldsymbol{V}\boldsymbol{P}^{(n-1)}_q\boldsymbol{V}^{T}+\boldsymbol{I}\right)^ {-1}\boldsymbol{y}_{m},
\label{alpha_q}
\end{equation}
\noindent
where $\boldsymbol{P}^{(n-1)}_q=$ diag$(\boldsymbol{p}^{n-1}_q)$ comes from the $(n-1)-$th iteration. The optimization procedure can terminate after a fixed number of iterations or when the following convergence criterion is satisfied:
\begin{equation}
\frac{ \left\|\boldsymbol{\alpha}^{(n)}_{m,q}-\boldsymbol{\alpha}^{(n-1)}_{m,q}\right\|_2} {\left\|\boldsymbol{\alpha}^{(n)}_{m,q}\right\|_2} < \Delta,
\label{SLIM_convergence}
\end{equation}
\noindent
with $\Delta$ a suitable small positive number. 
As for the SLIM initialization, we use the maximum likelihood estimate of $\alpha_{m,n}$ assuming that $\boldsymbol{z}_m=\boldsymbol{\sigma}_n \alpha_{m,n}+\boldsymbol{n}_m$ \cite{5654598,8781902}, namely
\begin{equation}
{\alpha}^{(0)}_{m,n}=\frac{\boldsymbol{\sigma}^{T}_n \big ( \boldsymbol{M}_m \big )^{-1} \boldsymbol{y}_m}{\boldsymbol{\sigma}^{T}_n\big ( \boldsymbol{M}_m \big )^{-1}\boldsymbol{\sigma}_n},
\label{SLIM_initialization}
\end{equation}
\noindent
for $n=1, 2,\ldots, N$ and $\boldsymbol{\sigma}_n$ is the $n$th column of $\boldsymbol{S}$. To automate the estimation of $q$, the Bayesian Information Criterion (BIC) \cite{1311138}, a model order selection rule, can be applied as in \cite{5654598}.

\section{Experiments}

\begin{figure}[!t]
	\centering
	\includegraphics[width=2.65in]{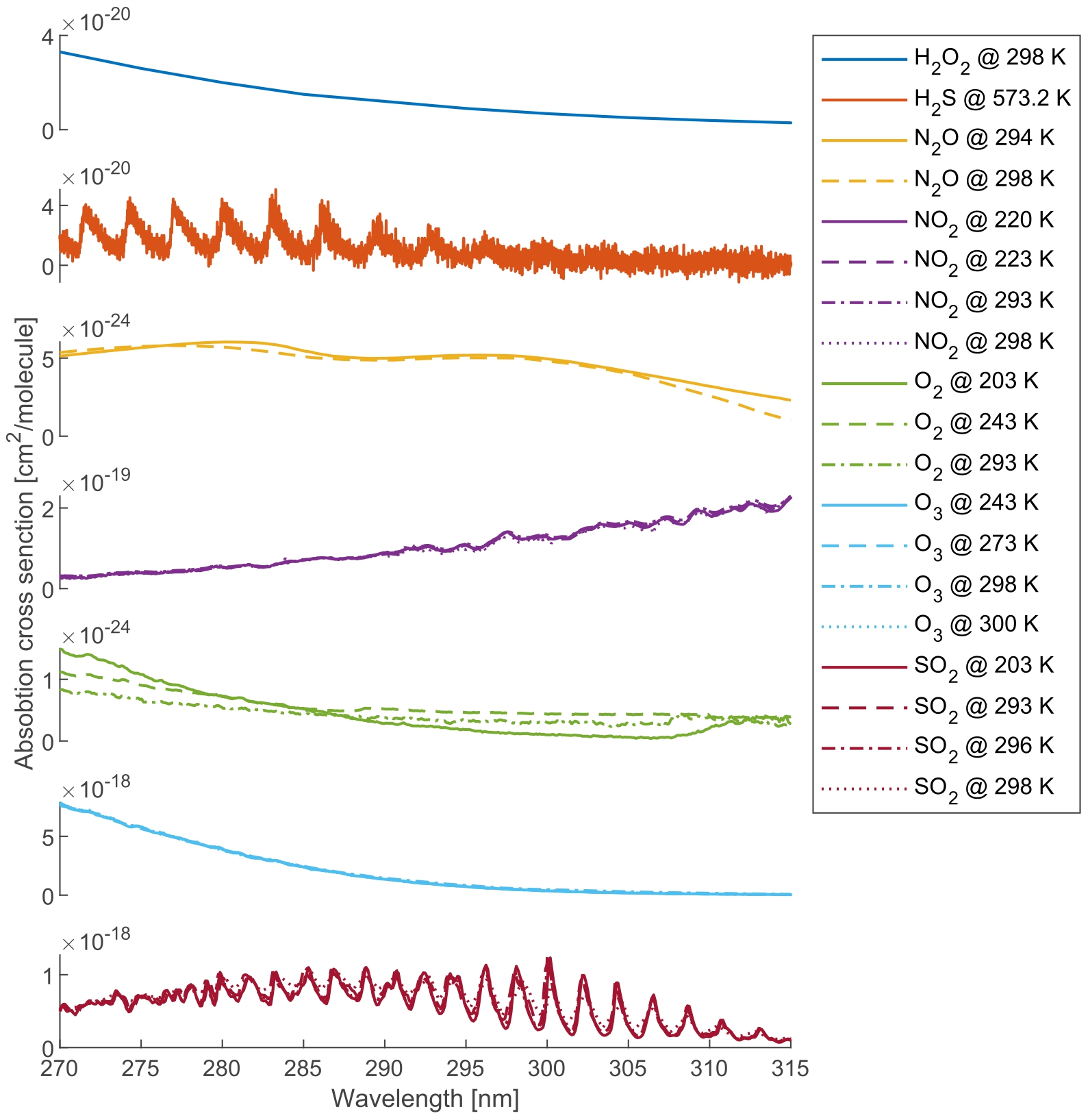}
	\caption{UV trace-gases cross sections (endmembers) used build up the dictionary of the considered sparse estimation problem. Species as well as their spectral regions are extracted from 
	\cite{essd-5-365-2013}.}
	\label{fig:Spectral_library}
\end{figure}
The spectral library is defined considering the main absorption gases in the UV region and it is composed by the cross sections extracted from the UV/VIS Spectral Atlas of Gaseous Molecules of Atmospheric Interest \cite{essd-5-365-2013}, which is a large collection of absorption cross sections and quantum yields in the middle UV wavelength region for gaseous molecules and radicals of primarily interest. Specifically, 
we consider the following main absorption gases at different temperatures to generate a consistent spectral library: hydrogen peroxide (H\textsubscript{2}O\textsubscript{2}) at 298 K, hydrogen sulfide (H\textsubscript{2}S) at 294.8 K, 423.2 K and, 573.2 K, nitrous oxide (N\textsubscript{2}O) at 294 K, 298 K and, 1428 K, NO\textsubscript{2} at 220 K, 223 K, 233 K, 265 K, 293 K, 298 K and, 300K, oxygen (O\textsubscript{2}) at 203 K, 243 K and, 293 K, O\textsubscript{3} at 228 K, 243 K, 273 K, 293 K, 298 K, 300 K and, 720 K, SO\textsubscript{2} at 203 K, 293 K, 296 K and, 298 K. It is important to note that, in the selection process, we take into account the absorption cross section for a specific gas and to a specific temperature that have a significant number of samples in the spectral region of interest. Figure~\ref{fig:Spectral_library} shows the more representative absorption cross sections in the UV band $270-315$ nm.

In the next subsection, we assess the nominal behavior of the proposed estimator using simulated data whose statistical properties perfectly match with the design assumptions. Then, we apply the new approach to real recorded data also in comparison with existing algorithms.

\subsection{Results derived from simulated data}

\begin{figure*}[!t]
	\centering
	\subfloat[1000 MC trials at SNR $=$ 20 dB.]{\includegraphics[width=2in]{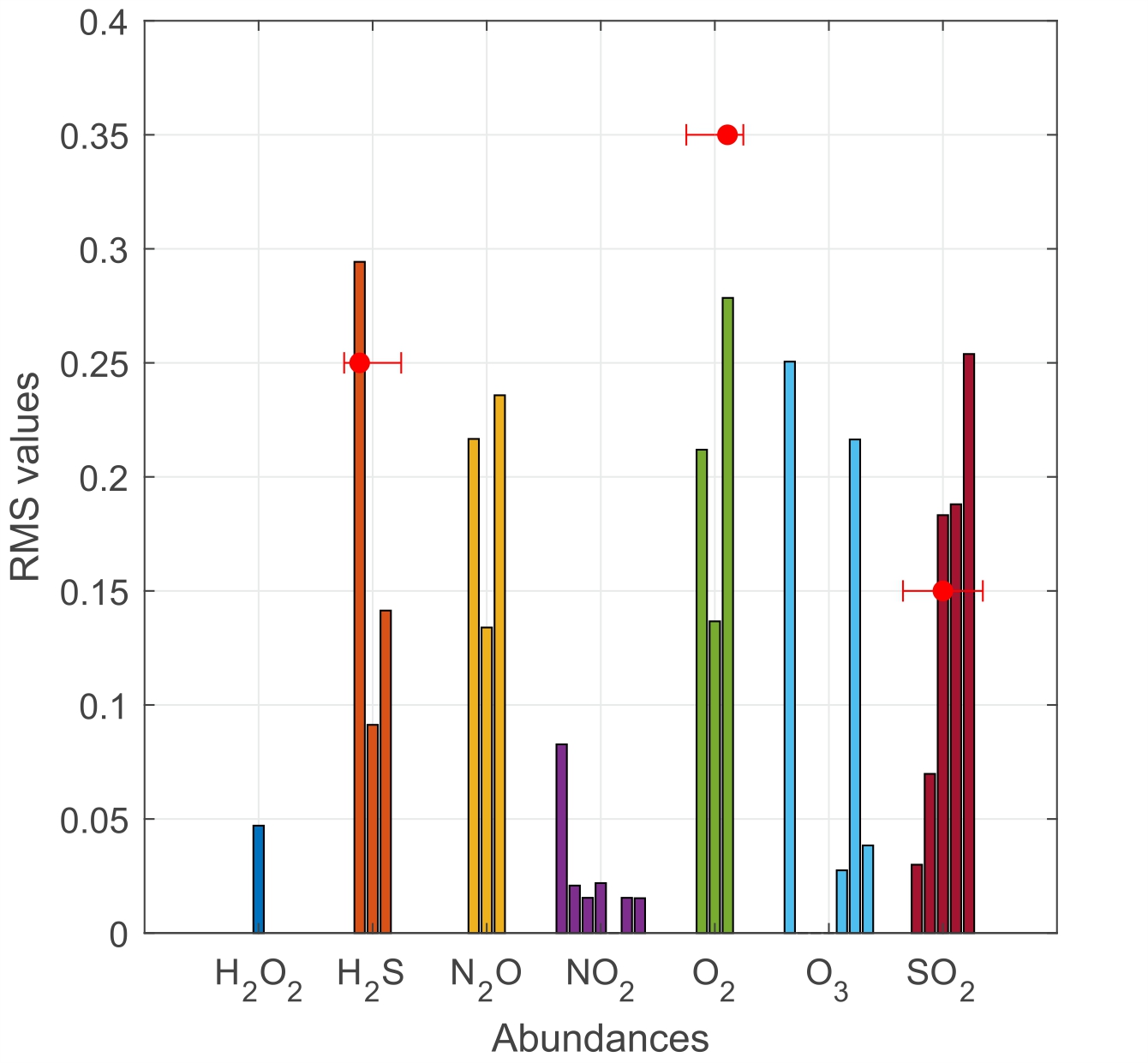}%
		\label{fig_fir_case}}
	\hfil
	\subfloat[1000 MC trials at SNR $=$ 40 dB.]{\includegraphics[width=2in]{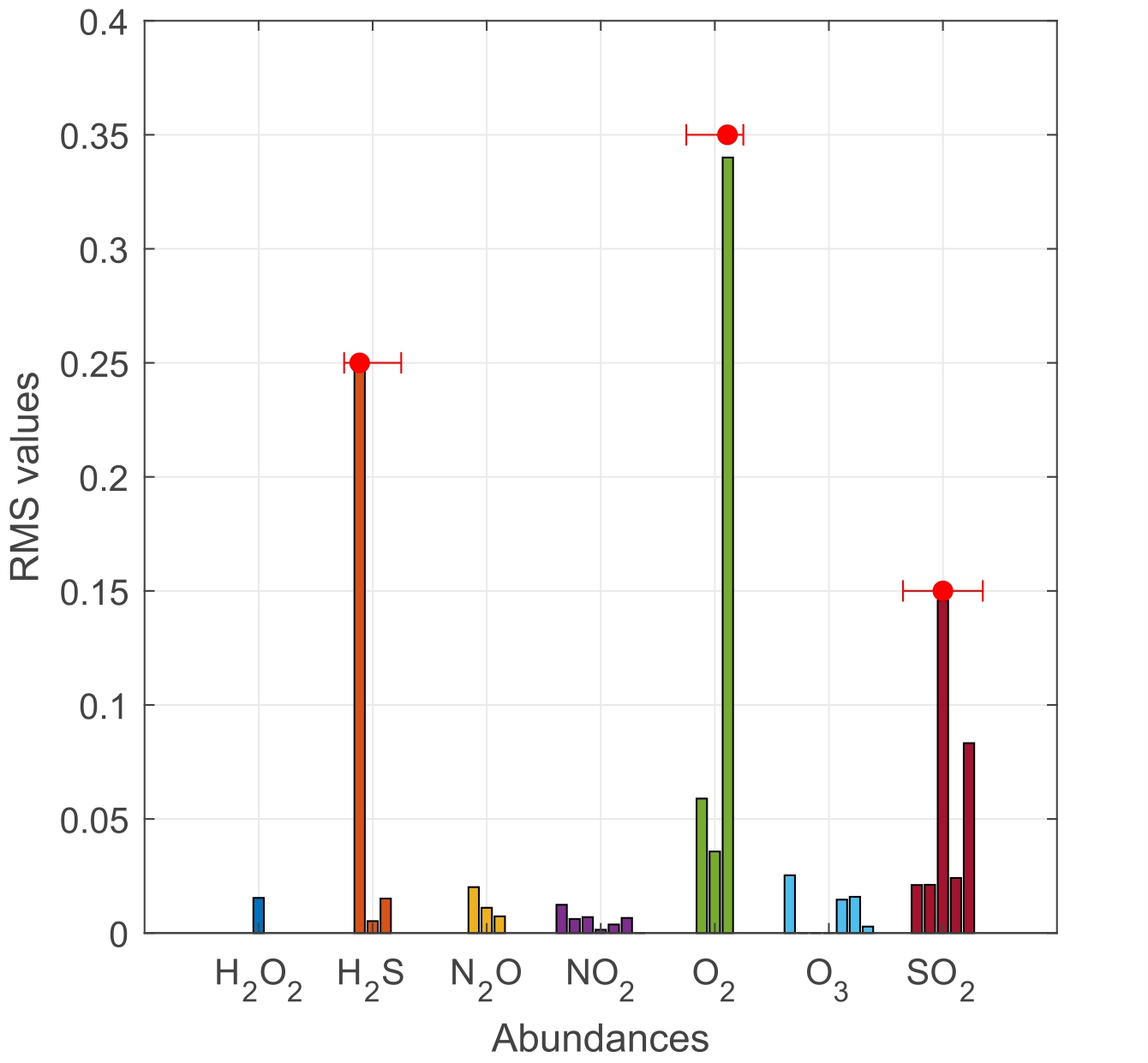}%
		\label{fig_sec_case}}
	\hfil
	\subfloat[1000 MC trials at SNR $=$ 60 dB.]{\includegraphics[width=2in]{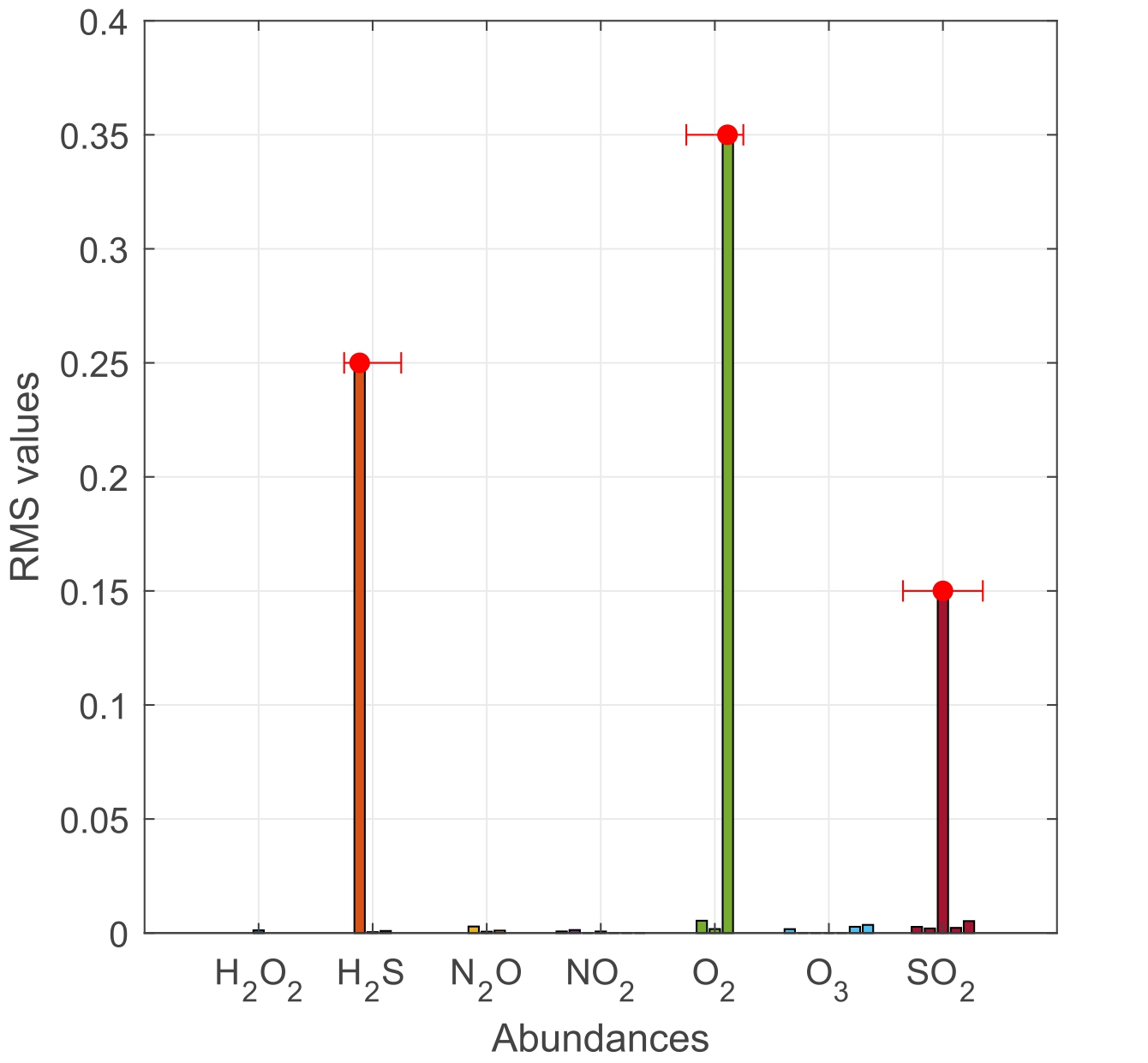}%
		\label{fig_thi_case}}
	\caption{Performance analysis using Monte Carlo method for simulated data. Analysis are performed at fixed SNR to evaluate the RMS value of estimated abundances. The segment amplitude is indicative of the number of specific gas available while the red dot corresponds to the ground truth.}
	\label{fig_sim}
\end{figure*}

\begin{figure}[!b]
	\centering
	\includegraphics[width=2.4in]{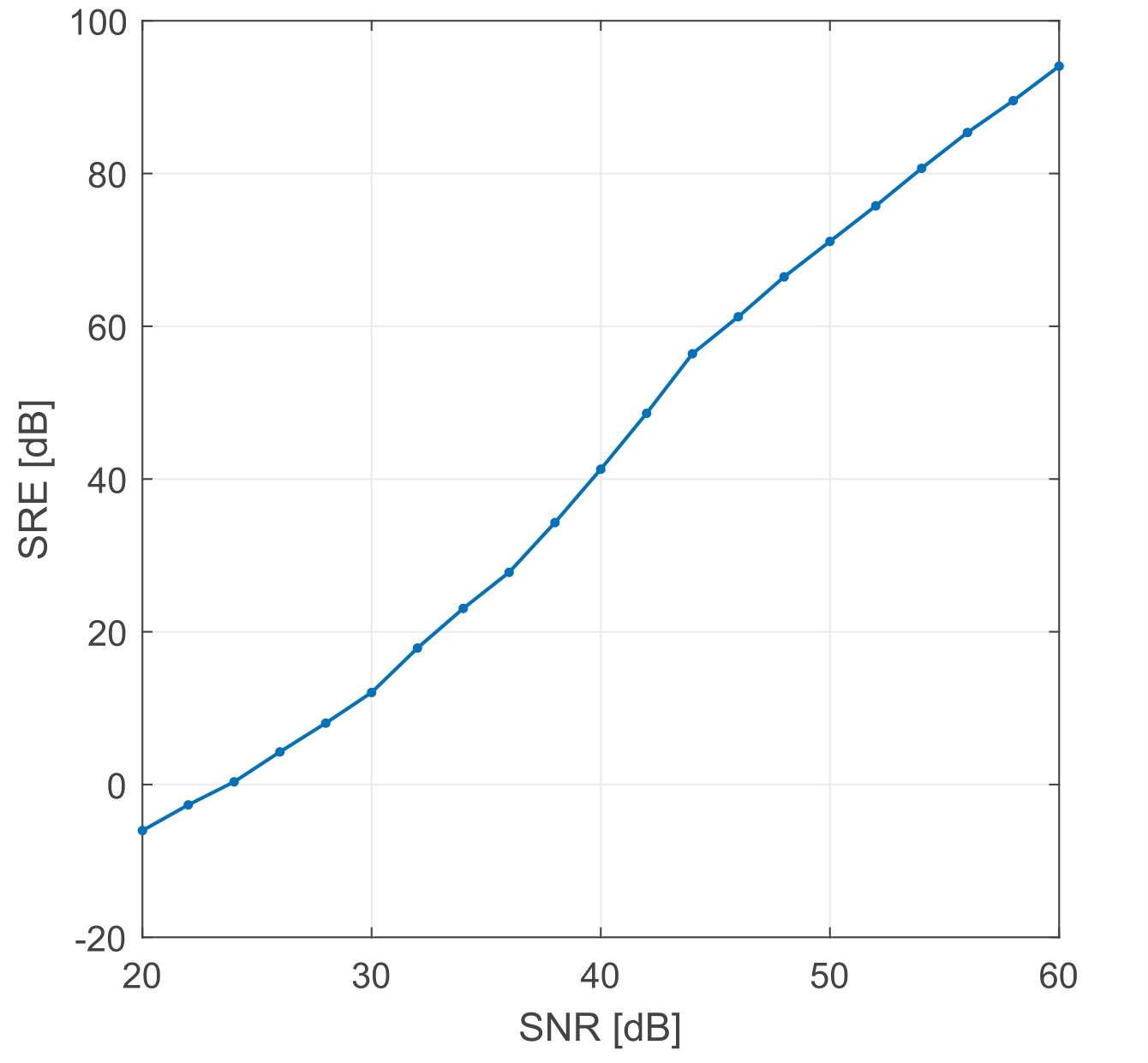}
	\caption{Performance analysis using MC method for simulated data.}
	\label{fig_sim_MC}
\end{figure}

The analysis on synthetic data considers a single pixel, i.e., $M=1$, with the presence of only three selected absorption gases within the whole spectral library: H\textsubscript{2}S at 294.8 K, O\textsubscript{2} at 293 K and, SO\textsubscript{2} at 293 K with a percentage of 25\%, 35\% and, 15\% respectively. For these gases, we consider $N=29$ absorption cross sections and $L=10$ spectral bands opportunely distributed \cite{1624604} in the interval $270 - 315$ nm. Thus the spectral library is $\boldsymbol{S}\in\mathbb{R}^{10 \times 29}$. In the simulated pixel, we assume the presence of gaussian noise with a noise power, $\sigma_n^2$ say, set according to the Signal-to-Noise Ratio (SNR) defined as
\begin{equation}
\text{SNR}= \frac{\left\|\boldsymbol{S} \boldsymbol{\alpha}\right\|_2}{\sigma_n^2},
\label{SNR_definition}
\end{equation} 
where $\boldsymbol{\alpha}$, is the true abundances' vector. 

The adopted performance metric is Signal-to-Reconstruction Error (SRE) defined as
\begin{equation}
\text{SRE}= \frac{\left\|\boldsymbol{\alpha}\right\|_2}{\left\|\boldsymbol{\alpha} - \boldsymbol{\hat{\alpha}}\right\|_2},
\label{SNR_definition}
\end{equation}
where $\boldsymbol{\hat{\alpha}}$ is the estimated abundances' vector using the proposed method. Such metric measures the quality of the reconstruction of the spectral mixture.

Figure~\ref{fig_sim} shows the obtained results. Specifically, Figure~\ref{fig_fir_case}, Figure~\ref{fig_sec_case}, and Figure~\ref{fig_thi_case} show the Root Mean Square (RMS) values of the estimated abundances calculated over 1000 Monte Carlo (MC) indipendent trials at the given SNR values 20, 40, and 60 dB. As expected, as the SNR increases, the RMS values approach the true abundances. Figure~\ref{fig_sim_MC} shows the SRE, expressed in dB, versus the SNR from 20 to 60 dB. From the inspection of the figure, it is confirmed that we achieve good performance for high SNR values.

\subsection{Results derived from hyperspectral satellite data}
The unmixing procedure has been applied to the retrieval of sulfur dioxide SO\textsubscript{2} volcanic emission using real data from the TROPOspheric Monitoring Instrument (TROPOMI), the single payload on board the ESA Copernicus Sentinel-5 Precursor (S-5P) satellite. 
S-5P is a gap-filler between the end of the OMI \cite{OMI} and the SCIAMACHY \cite{SCIAMACHY} instruments and a preparatory programme covering products and applications for Sentinel-5 mission.
The TROPOMI instrument is a nadir-viewing imaging spectrometer covering wavelength bands in the UV, VIS, near-infrared, and shortwave infrared wavelengths. 
The full spectral coverage of the TROPOMI UV band is $267-332$ nm with a spectral resolution of $0.45-0.5$ nm; the UV1 band ($267-300$ nm) has a spatial resolution of $5.5 \times 28$ km\textsuperscript{2} while the UV2 band ($300-332$ nm) is characterized by a spatial resolution of $5.5 \times 3.5$ km\textsuperscript{2} \cite{S5P_prod}. 

\begin{figure*}[!t]
	\centering
	\subfloat[SLIM based SO\textsubscript{2} concentration map.]{\includegraphics[width=2in]{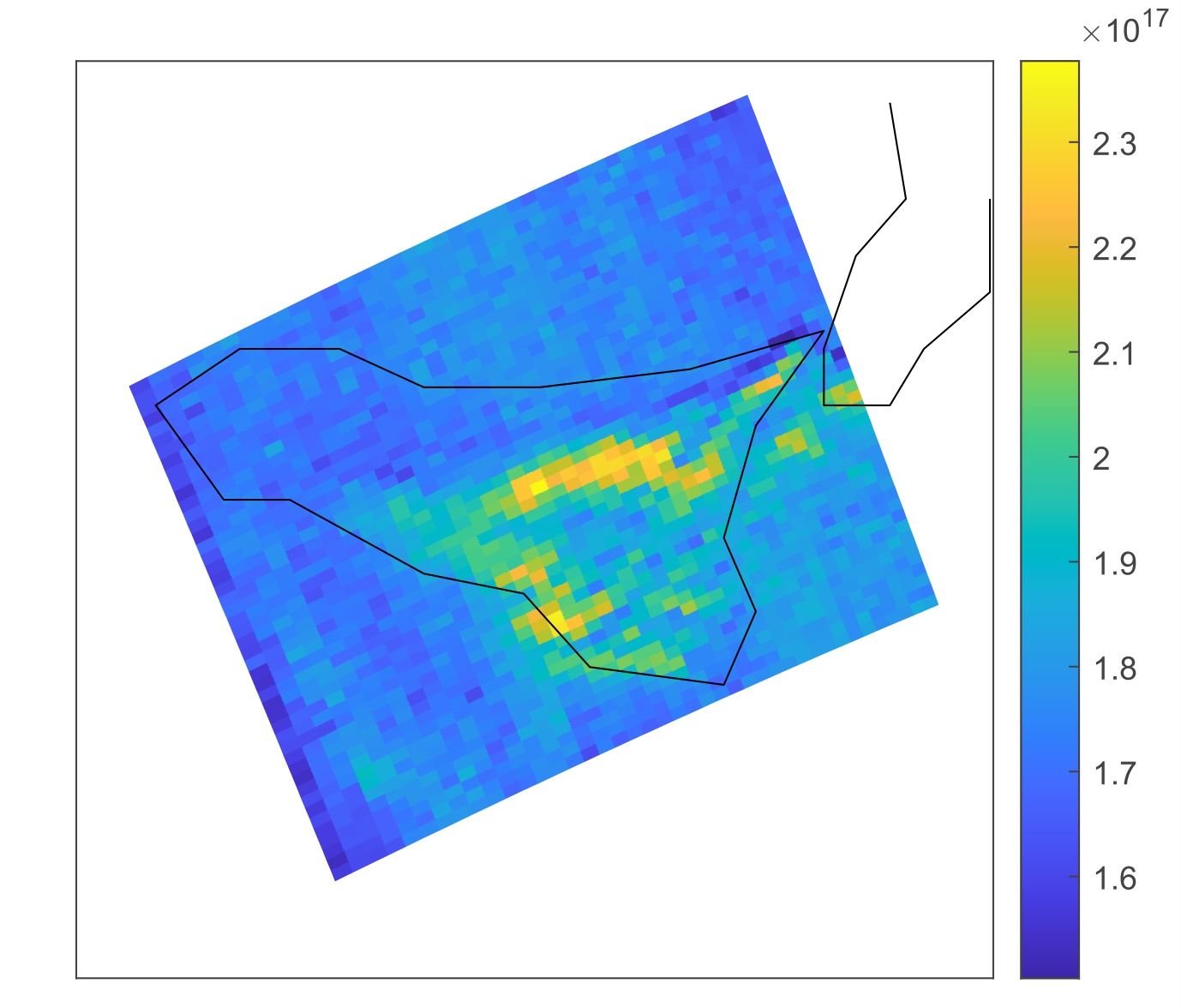}%
		\label{fig_SLIM_product}}
	\hfil
	\subfloat[BSS based SO\textsubscript{2} concentration map.]{\includegraphics[width=2in]{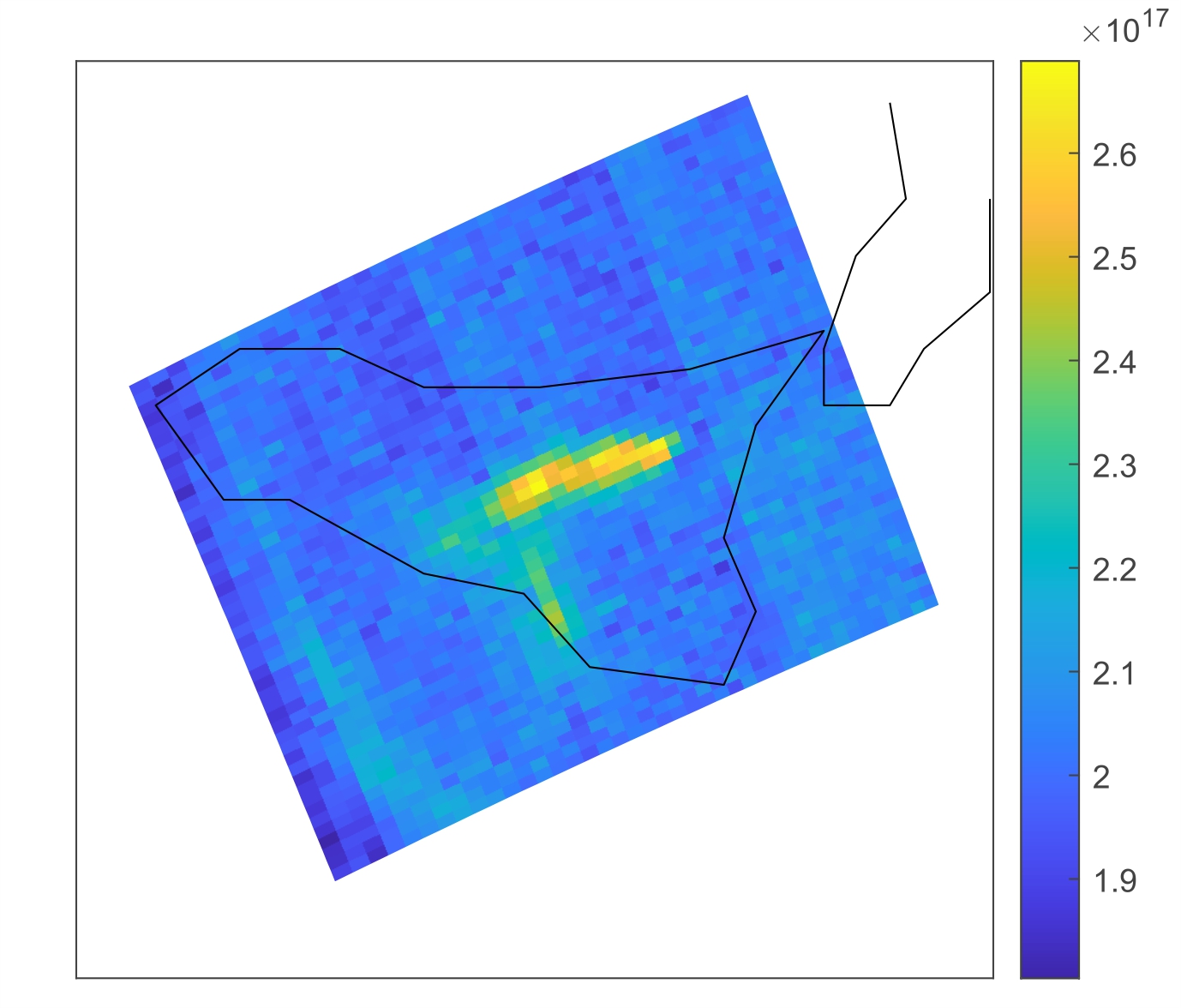}%
		\label{fig_competitor}}
	\hfil
	\subfloat[S-5P product for SO\textsubscript{2} concentration map.]{\includegraphics[width=2in]{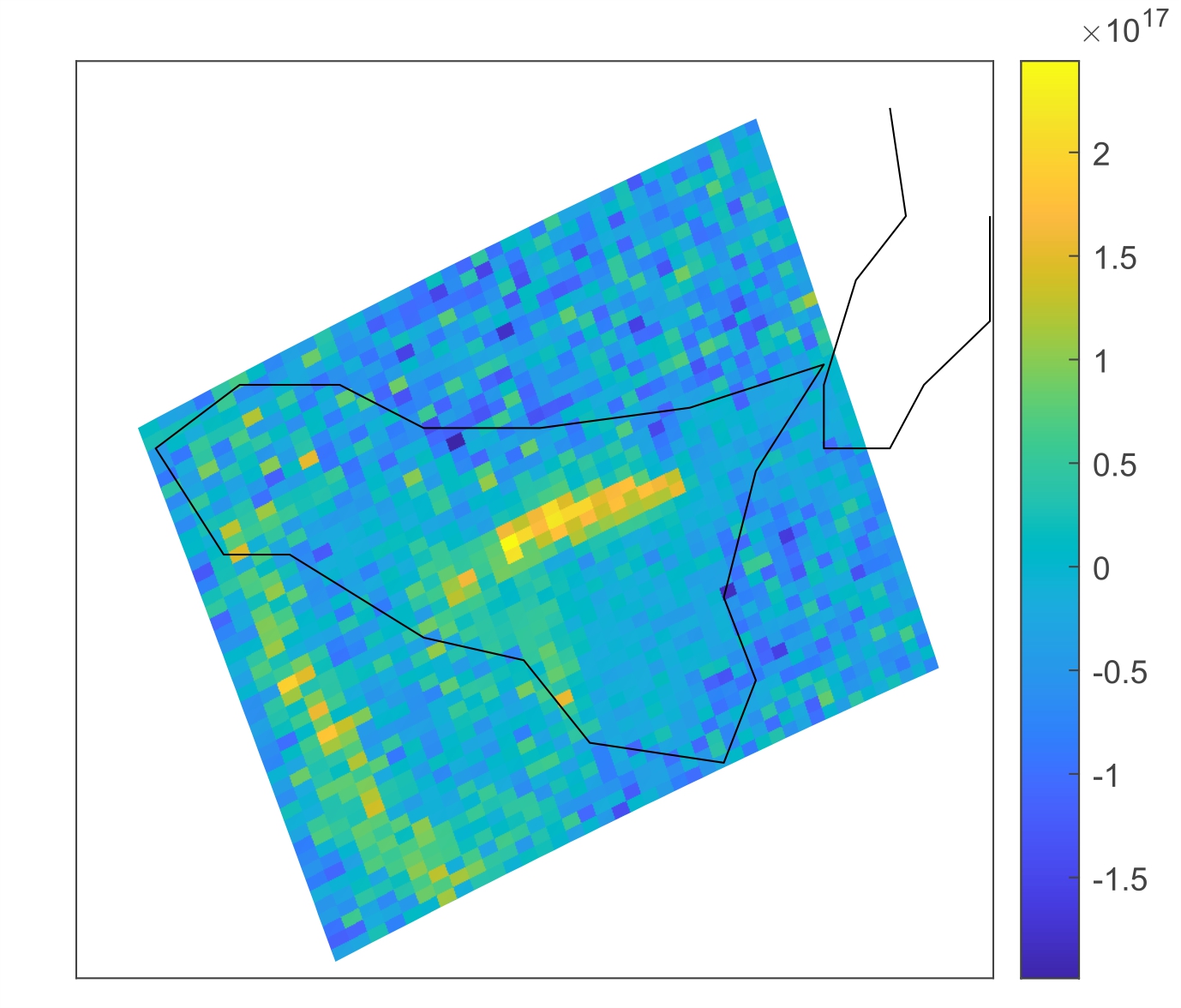}%
		\label{fig_S5_product}}
\caption{SO\textsubscript{2} total vertical column concentration map in molecule/cm\textsuperscript{2} over Sicily island from Etna volcano eruption on 22\textsuperscript{nd} of February 2021.}
\label{fig_product}
\end{figure*}

Data processed here are related to Italy’s Mount Etna, one of the world’s most active volcanoes, that has erupted in the middle of February 2021 spewing a fountain of lava and ash into the sky. The data, freely downloadable through the Copernicus Open Access Hub\cite{SCIHUB}, is captured on 22\textsuperscript{nd} of February 2021 during the ascending orbit number 17423.
As for all the Sentinel missions, the Sentinel-5P products are freely available to user. Specifically, we exploit Level 1B radiance and irradiance products and Level 2 geophysical products \cite{S5P_prod}. 
Our objective is to generate the SO\textsubscript{2} map over Sicily, applying the proposed procedure and considering the spectral cross section library defined in the synthetic simulations. We consider the dataset at UV2 band because it has higher spatial resolution and we clip the data over the area of interest. Specifically, we process the scanlines from 3032 to 3078 and the ground pixels from 49 to 89 so that we have a $41 \times 47$ pixel matrix. For each pixel, we have 497 wavelength samples, but we take into account only a subset in the spectral band $312-326$ nm as also defined by the S-5P/TROPOMI SO\textsubscript{2} algorithm \cite{S5P_prod}. Since the Sun's irradiance and the solar zenith angle are available for the UV2 band, we can evaluate the spectral reflectance as defined in (\ref{linear_mixing_model}), so we can apply the proposed gas retrieval methodology based on SLIM. Figure \ref{fig_SLIM_product} shows the obtained result which is the SO\textsubscript{2} concentration map in molecules per cm$^2$ over Sicily island on 22\textsuperscript{nd} of February 2021. From this image an eastward extension of the sulfur dioxide plume is apparent.  

For comparison purposes, we also report the results obtained by means the unmixing approach developed in \cite{6071033,7163519,7326687}, as well as the TROPOMI SO\textsubscript{2} total column provided by ESA as S-5P Level 2 product. From the visual inspection of Figure \ref{fig_competitor} and Figure \ref{fig_S5_product}, it turns out the presence of SO\textsubscript{2} plume in both images with comparable concentration density as also obtained in Figure \ref{fig_SLIM_product}. Considering the SO\textsubscript{2} S5P product as reference and according to its data quality defined in \cite{S5P_prod}, it is important to point out that both gases retrieval methodologies based on SLIM and BSS are within the product specifications. Moreover, we compute the Root Means Square Error (RMSE) over the entire SO\textsubscript{2} map and we obtain for both a comparable value that is about 2 DU\footnote{A Dobson Unit (DU) equals to $2.69 \times 10^{16}$ molecules per cm\textsuperscript{2}.}.

\section{Conclusion}

The sparse unmixing technique described in this letter is based on the linear mixing model where the combination of a great number of pure spectral absorption cross sections, \textit{a priori} known and available in the spectral library, is used to construct a sparse mixing model. 
With this strategy, the abundances' estimation process no longer depends on the selection of few particular pure spectral signatures nor on the capacity of a certain endmember extraction algorithm to identify such selected pure signatures. 
Our experimental results, conducted with both simulated and real data, are very interesting. In fact, this work represents the first application of a compressive sensing technique to retrievals of the SO\textsubscript{2} concentration map from hyperspectral data collected by the ESA TROPOMI spectrometer. 
Referring to the dataset acquired by the TROPOMI instrument over Etna active volcano, we obtained the SO\textsubscript{2} concentration map directly comparable with the Level 2 product released by ESA. Specifically, both retieval strategied based on SLIM and BSS have a comparable RMSE, but the proposed methodology offers the advantage of an unsupervised selection of the best candidate trace gas present in the scene.


%
%

\ifCLASSOPTIONcaptionsoff
  \newpage
\fi



%

%
%

\bibliographystyle{IEEEtran}
\bibliography{my_bibliography}

%

\end{document}